# DiSCoKit: An Open-Source Toolkit for Deploying Live LLM Experiences in Survey Research

Jaime Banks
School of Information
Studies
Syracuse University
banks@syr.edu

Jon Stromer-Galley
School of Information
Studies
Syracuse University
jrstrome@syr.edu

Samiksha Singh
School of Information
Studies
Syracuse University
ssing116@syr.edu

Collin Capano
Open Source Program
Office
Syracuse University
cdcapano@syr.edu

## ABSTRACT

Advancing social-scientific research of human-AI interaction dynamics and outcomes often requires researchers to deliver experiences with live large-language models (LLMs) to participants through online survey platforms. However, technical and practical challenges (from logging chat data to manipulating AI behaviors for experimental designs) often inhibit survey-based deployment of AI stimuli. We developed DiSCoKit—an open-source toolkit for deploying live LLM experiences (e.g., ones based on models delivered through Microsoft Azure portal) through JavaScript-enabled survey platforms (e.g., Qualtrics). This paper introduces that toolkit, explaining its scientific impetus, describes its architecture and operation, as well as its deployment possibilities and limitations.

## 1 Introduction

Across disciplines—from human-computer interaction and user experience design to education and futures of work—social scientists are increasingly attending to the dynamics and effects of how people interact with forms of artificial intelligence (AI). One such form includes large-language models (LLMs) and the technologies built upon them, such as conversational interfaces, narrow task-specific tools, multimedia generators, and artificial companions. For instance, researchers are exploring how conversation with nonhuman personas can enhance empathy for the represented entity (Li et al., 2025), how human trust is impacted when artificial agents perform step-wise planning in human-machine teaming contexts (He et al., 2025), and how AI performing role archetypes may contribute to convergent and divergent thinking in creative tasks (Kumar et al., 2025). There are non-trivial challenges with conducting such research at scale, especially some barriers to deploying LLM-based stimuli through web-based interfaces in order to garner larger samples. We developed the <u>D</u>aemon for <u>i</u>n-<u>S</u>urvey <u>Co</u>nversation Tool<u>kit</u> (DiSCoKit) to mitigate these challenges. This framework is designed for deployment by researchers with limited technical expertise with support from standard university IT resources. In this paper, we discuss the scientific motivation for development the toolkit, its architecture, implementation best practices, capabilities and decisions to be made, and ethical and practical considerations.

## 2 Scientific Impetus: Scalable Online Deployment of Controlled AI Stimuli

Online surveys have long been leveraged in human-machine communication (HMC) and human-computer interaction (HCI) research because they are relatively low-cost, quick to deploy, and afford access to large or hard-to-reach populations (Müller et al., 2014). Although useful in these ways, using surveys in human-subjects studies has also been critiqued for an overreliance on reporting perceptions without capturing actual behavioral content or dynamics as humans interact with a focal piece of hardware or software (Williams et al., 2017). Notably, a practical issue at the center of that critique is the technical difficulty of deploying ecologically valid technology-interaction experiences *through* survey interfaces. These conditions often leave researchers in a challenging position: Choose between (a) deploying a live interaction experience in a laboratory setting (where behaviors can be captured through direct observation), with necessarily small samples or long data collection periods (as lab-based samples are constrained by space, time, and resources) or (b) deploying surveys to remote, often-unverifiable samples as they rely on recall, untrackable activities, non-interactive scenarios, and/or hypotheticals. In other words, we must make trade-offs between feasibility and validity.

Rather than adopting either set of limitations, we worked to minimize those trade-offs by developing a solution that permits the deployment of a live, loggable (i.e., full conversation capture), adaptable and believable, manipulable conversational-AI experience through a common online survey platform. Moreover, we aimed to go beyond standard deployments to explore how the solution might allow for manipulating the AI to *change* its behavior over the course of the interaction.

The specific challenges overcome and opportunities leveraged include those presented by LLM technologies, by the survey platform, and by the scientific work. First, due to their stochastic nature LLMs are unpredictable, often generating unexpected content whereas empirical studies are carefully designed and experiments require tight controls over stimuli. This unpredictability can, however, be managed through temperature parameters and system prompts (the master set of rules that an LLM should follow), and multiple system prompts can be transparently layered (see Neumann et al., 2025)—even injected at various stages in an interaction to shape a shifting behavior trajectory. Second, Qualtrics XM (2025) and Google Forms—among the platforms most widely adopted by research institutions—are relatively locked down, with a customization interface that permits specific types of content with specific properties. However, they *do* offer application programming interfaces (APIs) and affordances for integration with other systems, including webhooks that can act based on events and trigger external systems, a web service element that supports JSON elements, and embedded data that allow information to pass to and from other services. Third, social-scientific human-subjects research demands clear, error-free, believable human-AI interaction. Experimental designs add further requirements: AI behaviors must remain highly consistent across conditions—despite varied participant inputs—except for the specific variable being manipulated (see Shadish et al., 2002).

In developing a toolkit (i.e., a collection of customizable code resources) that balances these limitations and requirements, we aimed to formulate the solution using open-source elements where possible, so other researchers could leverage this solution to overcome the same barriers.

"Open Source" (OS) is a set of principles and criteria for the development, dissemination, and licensing of software, including free distribution, access to source code, permission over derived works, non-discrimination in access, technology neutrality, and non-restriction of other software (Open Source Initiative, 2024). For DiSCoKit, the OS component of the toolkit extends to the core app and the bridges among proprietary platforms (as both the focal survey platform and various LLMs are proprietary technologies). As part of Open Research more broadly (McKiernan et al., 2016), OS software can help improve transparency, community dialogue, new ways of thinking about data, and adjacent knowledge production (von Krogh & Spaeth, 2007). It can also contribute to a freer exchange of ideas and replicability of results, with science centered as a public good (see Willinsky, 2005). With this publication and accompanying repository, we aim to support these ideals in relation to the study of human-machine interaction

## 3    Toolkit Architecture

DiSCoKit operates through a multi-layered architecture that bridges the gap between proprietary survey platforms and LLM

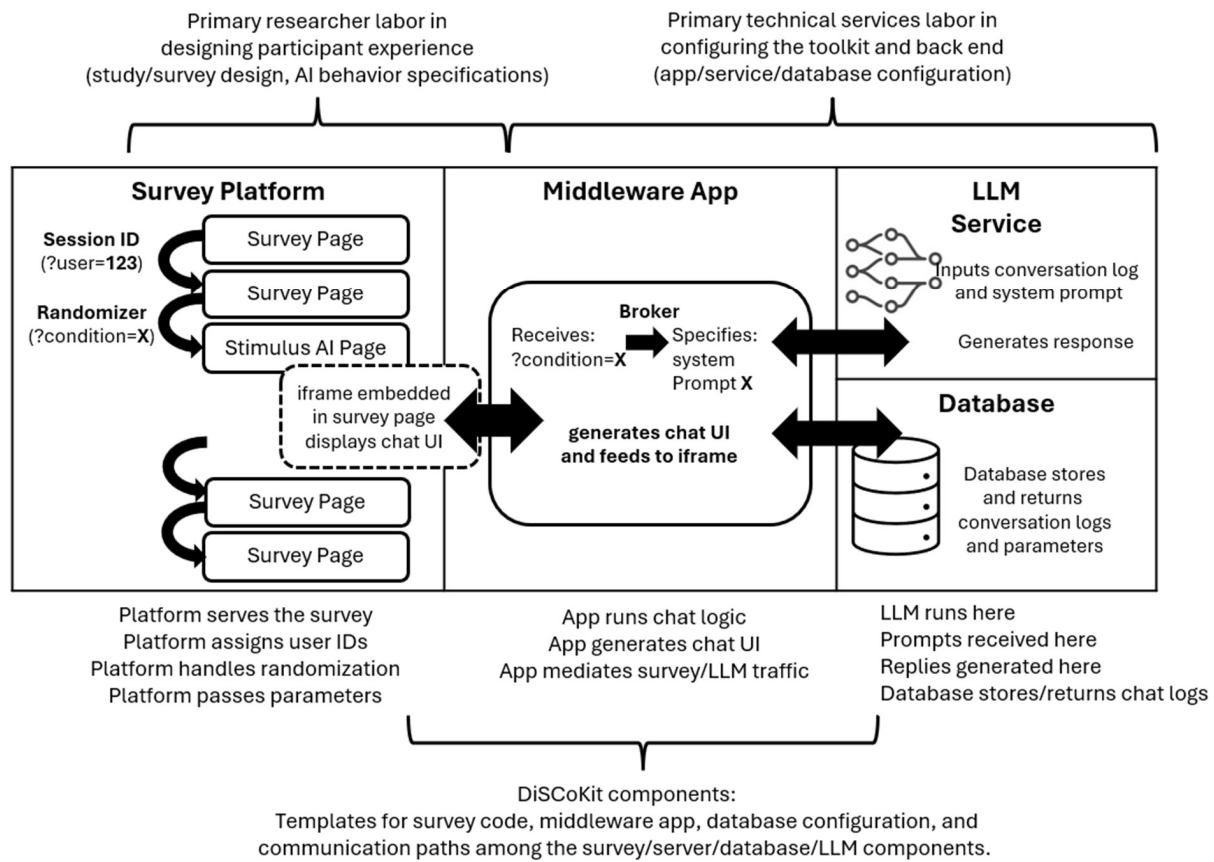

**Figure 1: Diagram of the DiSCoKit Architecture**

services. In doing so, it provides persistent storage for research data, maintains flexibility required for naturalistic human-AI interactions, and allows for control required of various study designs. As a daemon-like service running in the background of a live survey session, it is transparent to users, brokers data flows among elements in the system, and customizes that brokerage based on events in the survey workflow.

## 3.1 Core Components

DiSCoKit comprises three key components working in concert (Figure 1): The primary (1) middleware app that brokers data and workflows among other components, (2) code embedded in the survey platform that links it to the app, and (3) connections to a database and to and LLM service that serve and store the conversations. This architecture emerged from the practical constraints and opportunities presented by conducting controlled experiments with live AI systems: The need to manipulate AI behaviors across conditions, ensure participant privacy and session isolation, track experimental assignments, and capture behavioral data alongside self-report measures.

The **survey platform** delivers survey content to human subjects via a web interface. For the toolkit template, we relied on Qualtrics XM survey platform. Leveraging a block of simple JavaScript (code commonly used to manage web site behaviors), this component brings the chat interface directly into the survey window so participants can easily have a natural conversation with the AI within the survey web interface—much the same way that models become accessible to users through ChatGPT, Gemini, or Claude web interfaces. When a participant lands on a survey page with the embedded chat interface, JavaScript covertly sends survey session details (e.g., participant's survey-session ID and assigned experimental condition) to the app, facilitating the actual interaction and allowing other components to track the interaction data.

The **middleware app** manages the human-AI interaction. This component is powered by the toolkit app based on Flask 3 (Pallets, 2025) and is activated when the survey platform sends a message to the middleware—that is, a request to display the chat interface for a given participant and a given behavior or experimental condition. In response, the middleware builds an interactive experience governed by a multilayered system prompt and selected LLM service according to that request. Once a response from the LLM has been returned to the middleware, it may pass the response directly to the front-end chat interface, augment the LLM response, or potentially make additional LLM requests to further refine an appropriate LLM response in line with the desired stimulus experience.

Finally, the middleware is linked to two back-end resources. First, the app is linked to the **LLM service**. The toolkit is templated for use with serverless LLM instances hosted by and configured through Microsoft Azure, however the middleware can be extended to other LLM providers including session-based LLM providers. Importantly for the toolkit design, Azure's serverless LLM facilities are "stateless" in that any prior conversation in a session is unavailable to the LLM unless the middleware actively provides that context to the LLM as part of a new round of dialogue. While the DiSCoKit system technically supports "stateful" LLMs that require management of state through JavaScript, the stateless nature of the Azure endpoints allows DiSCoKit a great deal of flexibility with respect to guiding the chat interaction. For example, it is possible to change the system prompt mid-conversation to govern how the model behaves over the course of an interaction. This design choice permits fine-grained control over the conversational flow, as each survey page advance can trigger new prompt injections that modify AI behavior mid-conversation. Stateless systems have the additional benefit of being more secure and more scalable than systems that must maintain client state.

The middleware is also linked to a **database** keeps a detailed record of every turn and whole conversation, linking each message to the participant's survey-session ID, their experimental condition, and an event timestamp. The database supports reporting of this data at several levels of aggregation, helping researchers view results in a spreadsheet or tie results to data extracted from the survey. For example, a researcher can generate CSV datasets where each row represents a statement made by either the participant or the LLM or alternatively one where each row represents the entire conversation between the participant and the LLM. This logging is essential because it permits downstream data checks to determine whether the live AI behaved as intended, retains whole conversations for analysis, permits connecting chat data with survey answers using the anonymous ID, and indicates which components of the system prompt are active at any point in the conversation.

Of note, toolkit configuration should be driven by the scientific question at hand, so the researcher does coordinate with technical team members to specify some parameters, including variable naming conventions, the number and types of any experimental conditions, AI behavioral allowances and constraints, and other critical study characteristics that should be translated to connective code or AI system prompts.

## 3.2 Data Workflows

To illustrate how the three components work together to support workflows that enable participants to experience the live LLM interaction, it is useful to follow the story of a generic user experience. Aligned with the template, this flow will rely on the Qualtrics survey platform. From a participation announcement or invitation, a participant clicks a survey link that loads the first page of the Qualtrics survey; at the start of that session, Qualtrics generates a random alphanumeric participant ID (e.g., P_a#4567y). The participant completes the informed consent procedures as they would in any other survey. In this exemplar experimental design, Qualtrics' randomization function assigns them to one of the available experimental conditions (let's say a condition labeled "3"). Qualtrics maintains these pieces of metadata for the duration of the participant's survey session. When the participant reaches the page in which the chat experience is to appear, Qualtrics passes this metadata to the DiSCoKit middleware app. Based on that request, the middleware then establishes the condition-specific chat behaviors (behaviors associated with condition 3, for example) governed by a condition-specific system prompt and selected LLM. The middleware then constructs the chat user interface inside the Qualtrics page to support interactivity. The metadata specifying the experimental condition or specific behaviors are stored in a JSON file that can be managed outside of the DiSCoKit code context. In theory, there are no limits to the number of experimental conditions or behavioral variations that could be defined. Once the specific LLM behavior is called and the chat interface is rendered, the DiSCoKit system allows conversation to flow naturally, with each turn logged to the database with timestamps, the participant ID, condition identifier, and the content of both the human and AI messages. At any point in the exchange, Qualtrics' page-advance feature can be leveraged to move to a new survey page, and the toolkit code in that new page delivers a new request to re-display the chat in the survey. In that request, researchers can also deliver a middleware call for a new system prompt to change the behavior of the LLM. For instance, a page advance could call a system prompt that makes the AI change its demeanor, language style, or competencies, otherwise perform controlled behaviors as required by the study design. Multiple page-advances and prompt-calls can be strung together to progressively change behavior over time.

### 3.3   Survey Configuration and Communication

Researchers must embed a block of JavaScript within the survey itself; that embedded code permits the survey platform and middleware app to communicate so the LLM chat window is rendered and managed within the survey itself. Within Qualtrics, this code is most easily added to a "Text/Graphic" question item through the HTML editing view, or (for more control) through the "Add JavaScript" option for extra customization. Adding the script to a Text/Graphic item allows design flexibility since moving the item also moves the chat window; it isolates the window to that item ensuring the chat functions and display will not interfere with other survey elements. The researcher-configured code specifies the size and location of the chat window (iframe), and how metadata such as participant ID is passed to the DiSCoKit middleware.

Survey platforms often assign participant IDs; in Qualtrics, this ID can be stored as embedded data, which allows values to be passed through the survey flow. At the start of the survey, two pieces of embedded data are typically established: The participant's unique ID (generated automatically by the survey platform or sometimes predetermined by researchers) and the experimental condition assignment determined by the survey's random assignment feature (or some other assignment function). Note that if the study design is *not* an experiment, the condition assignment is of course not relevant, since all participants experience the same AI behavior.

All DiSCoKit logging (i.e., the capture and persistence of rounds of chat and chat parameters) happens in parallel with participant interaction and is invisible to them. Even if a curious survey-taker tried to tinker with the browser, their actions would not affect the secure and accurate record of what actually happened in the chat exchange. This separation of duties between components–the survey platform's management of survey logic, randomization, and interfacing; the middleware app's communication with and management of the LLM; the LLM's generation of conversation text and database's logging interaction events—helps to both flexibly satisfy scientific requirements and robustly address common technical barriers such as anonymous persistence of activity and mitigation of security issues.

## 4   Installation and Configuration

DiSCoKit requires a minimum suite of technical assets to be functional. These are assets that should be accessible to most universities via central IT infrastructures or unit-specific resources: Infrastructural assets, LLM access, and a survey platform license. Regarding infrastructure, the system relies on a Linux server or Docker environment supporting Python 3.12. DiSCoKit middleware requires an LLM for chat inference; the toolkit template[1] is based on Azure OpenAI since it offered the simplest integration. Serverless endpoints in Azure support many Open AI models and by extension DiSCoKit supports them as well. Of note, the toolkit can be tailored to work with other LLMs. For the survey platform, template JavaScript is included that supports the inclusion of a DiSCoKit conversation in the context of a Qualtrics survey, since that platform is commonly used by investigators at research institutions. The included JavaScript can be used as a reference to include a DiSCoKit deployment on any web survey platform that supports JavaScript and can make HTTPS requests.

DiSCoKit can be configured and deployed by non-technical research teams who have access to basic institutional IT support. The process involves preparing the server, set up the Python environment, configuring secret files (i.e., account logins), and configuring study settings (e.g., conditions and condition-specific system prompts). Once the DiSCoKit middleware is operational, the chat interface can be integrated with the survey platform through typical survey-design processes (e.g., defining a survey flow according to study requirements like experimental condition assignment) and then adding the template JavaScript to embed the chat interface, customizing as required for the design. Details on these technical processes are available in the DiSCoKit repository hosted at https://github.com/SHASTAlink/DiSCoKit.

There are several important possibilities and limitations that researchers and supporting technical staff should be cognizant of, with implications for both scientific and technological processes and outcomes.

---

[1] Here, we use 'template' in a generic sense to refer to the default set of code resources that can be customized (rather than the Python-specific reference to behavioral design patterns).

## 4.1   Features and Capabilities

The system, as detailed by the documentation in the online repository, has a number of features that can support rigorous scientific efforts with both flexibility and control.

**Adaptation to different LLM and survey technologies.** Although our toolkit reference and documentation is tailored for the deployment of Azure-based ChatGPT through Qualtrics, it is feasible to adjust the reference code to accommodate other technologies. For example, since LLM behavior is dictated by parameters passed to the DiSCoKit middleware via a URL query string, almost any web page or survey-platform page that can host an HTML iframe should be able to host a DiSCoKit-based interaction. Azure's stateless LLM endpoints supports hundreds of alternative LLM backends. DiSCoKit can be easily configured to support them.

**Continuous chat sessions—consistent or changing.** The chat session can persist across survey pages within the same session—there can be one continuous, cohesive conversation carried across multiple sets of questions. This means, for instance, that participant attitudes can be captured at multiple points across a conversation, or different questions relevant to different conversation topics or stages can be presented. For further flexibility, each new page presents an opportunity for a new prompt injection—cascading prompts that add or change behaviors can be written into the DiSCoKit middleware layer and toggled on or off for each survey page. This capability can be combined with other survey features like automatic page timers to initiate controlled AI behaviors over the course of a survey experience.

## 4.2   Limitations and Open Questions

There are some limitations to this toolkit that should be considered, as well as uncertainties around the range of possible deployments.

**AI Behavioral Control and Bias.** System prompts are a generally effective way to guide LLM behavior, however, it should be acknowledged that they are not guaranteed controls and there may be variability in the AI's behavior from session to session depending on the LLM used and its settings. There may also be tensions between system prompts and an LLM's inherent guardrails or training, and creative prompting may be necessary to accomplish the necessary behavior. This is an inherent trade-off: Accepting variability in chat behaviors in exchange for a live, more ecologically valid human-AI interaction experience. This trade-off could be mitigated by tailoring the app handshake to coordinate with a locally operated LLM allowing for more control. Beyond managing that tradeoff, extensive iteration and testing is required to craft a prompt (and derived behaviors) that are scientifically defensible in relation to specific study goals, planned tests, and theoretical frameworks. A well-written and -vetted prompt strikes a balance of control (corralling behavior inside a specific scope and operationalizing focal variables) and the adaptability and believability of a live LLM experience (versus more trite and rigid bots or rule-based interactions).

**Privacy and Security Considerations.** A note on privacy and anonymization is warranted here, given the potential sensitivity of conversational data and the privacy-protection requirements of human-subjects research. Human-subjects research requires measures to protect participant privacy. In relying on pseudonymous participant IDs for matching survey responses to logged chat data, there is some measure of privacy protection. However researchers should still employ best practices around the potential for identifying information in the chat to be attributable to any identifying information in the survey panel or responses. Care must also be taken in selecting survey platforms and servings for deploying the DiSCoKit system. For our deployment and the template configuration, we rely on the university-licensed Qualtrics, university-licensed Microsoft Azure (an enterprise license that ensures data is not used to train the model), and university-controlled server as measures to maintain control over data storage and access. Regarding system security, sensitive information like API keys and database details is kept in a private settings file on the server.

**Concurrency Challenges.** Studies expecting very high concurrency (i.e., those with hundreds of people participating simultaneously), might consider alternative database providers. The template configuration uses SQLite, as it is simple and adequate for most DiSCoKit deployments. For most research with staggered participant recruitment, this is sufficient. Studies with burst patterns in traffic may benefit from databases with integrated and dedicated concurrency management (e.g., PostgreSQL). Fortunately, the database interface to DiSCoKit is abstracted by Flask so shifting among databases, if required, does not require a complete rewriting of the app.

**Public-Facing Servers.** The toolkit template implementation of DiSCoKit runs on Linux and Python 3.12. Other platforms might work correctly but have not been tested. It is likely that other, tailored implementations will need to be "visible" to the internet; in the deployment example detailed below, a public-facing server was used alongside a platform for managing the containerized app. An exception would be cases in which participants are all using systems within an institution's internal network. In that case, DiSCoKit does not need to be visible to the wider internet even when being hosted by a web-accessible survey platform.

**Possible LLM Services.** The current toolkit is predicted on stateless LLM services, but customization of the toolkit would allow for integration of stateful LLMs. This potential would, for instance, support integration with retrieval-augmented generation (RAG) systems and vector databases.

## 5   Sample Deployment Case: 3x3 Experiments with a Forced AI Mistake

DiSCoKit was originally developed for an experiment manipulating priming language (three levels of anthropomorphic language in an LLM description) and visual cues (three different types of icons attributed to each turn of the LLM in the chat). Requirements of the study included the ability to vary the AI icon appearance based on a randomly assigned condition in Qualtrics, to log the chat, and to be able to match up the chatlog with the survey responses—and these requirements drove much of the toolkit's feature set. Additionally, the context for the study drove a particular *kind* of interaction—one in which the AI would make a mistake. Based on prior studies, we elected to focus the interaction on a task that was relatively short (< 5 minutes), easily accomplished without technical knowledge, afforded an easy identification of the mistake, and that was tightly scoped to minimize variability across participants. Specifically, we had them write an English-language poem of 8 lines that would rhyme in an AABB scheme—but the AI would *always* make an error in one of the four line-pairs by injecting words or phrases written in Japanese script.

In this implementation, the participant would start the survey and complete informed consent procedures. Then, each participant was randomly assigned to one of the three language-prime conditions and answered questions about the prime; they then proceeded to the instructions for the task. After that, participants were randomly assigned to an icon condition and directed to a survey page designated for that condition—up to this point, all through native Qualtrics functions. On that condition-specific survey page, the standard text heading was augmented with the JavaScript code that embedded the iframe by passing survey metadata to the DiSCoKit middleware and by extension, displaying the LLM chat application on page load, DiSCoKit determined the relevant system prompt, and continued to serve the chat session in the frame with the corresponding visuals and behaviors.

Each condition-specific session-deployment page featured a five-minute countdown timer visible to the participant (a native Qualtrics feature). When they were informed that they had five minutes to complete the task. When the timer reached zero, the survey was configured to automatically advance to the next survey page. On that next page, additional metadata was specified when re-embedding DiSCoKit to instruct the system to (a) disable poem-writing behavior and (b) reject any further request to help with poem-writing, though it permitted other conversations. This constraint helped to maintain temporal consistency across participants (i.e., everyone worked on the poem for only five minutes with however many mistakes would manifest in that period) while still permitting the AI to believably function. That second page included a question asking the participant to report the task outcomes by copy-pasting the last generated version of the poem, and the active chat session allowed them to review the conversation for that reporting. Throughout 15 sessions of research-team trials and external pilot testing, the AI maintained the necessary error behavior for all but one session; in 218 sessions during live data collection it maintained that behavior for all but 16 sessions. Pilot testing also revealed that the initial model (ChatGPT 5mini) was found to be too slow—enough so that it made accomplishing the task in the five-minute window to be challenging. As a result we switched the model (via the Azure portal) to ChatGPT o4 and adjusted the system prompt to convey shorter answers and to avoid some repetition that was annoying pilot testers. That small adjustment resulted in faster (more believable) performance and a more task-suited persona for the model.

The DiSCoKit deployment was ultimately extremely flexible. Not only did the architecture permit the quick and simple tuning of the model and its behavior, it permitted replication of the study design with two additional variable sets—one study in which the AI-cue manipulation were different names (such that the condition identifiers cued up system prompts that generated a predetermined name that was more humanlike or machinelike or displaying no name at all) and another that manipulated the way the AI referenced itself (in first-person, third-person, or not at all). Those variations can also be seen in the above-linked app demo.

# 6  Sample Deployment Case: 3x3 Experiments with a Forced AI Mistake

DiSCoKit is a low-barrier, low-cost, scalable, customizable toolkit for deploying live LLM interaction experiences via online survey platforms. It empowers researchers to forego rigid, rule-based chatbots in favor of more believable and dynamic conversational agents, embedding them directly into the survey interface. AI behavior is reasonably controlled through vetted system prompts, which can be layered and toggled on or off based on simple scripts embedded in individual survey pages. Ultimately, this toolkit helps human-AI interaction researchers to conduct more ecologically valid, scalable, and valid studies while maintaining behavioral controls—even in multivariate experiments. We challenge the potential user community to adopt, customize, and extend this toolkit to even further advance the possible benefits to research in this domain.

## CODE AVAILABILITY

## CRediT STATEMENT

JB: Conceptualization, Writing, Editing, Supervision, Administration, Funding Acquisition. JSG: Software, Editing, Supervision. SS: Software, Writing. CC: Editing, Funding Acquisition.

## FUNDING

Research was sponsored by the Army Research Office and was accomplished under Grant Number W911NF-25-1-0079. The views and conclusions contained in this document are those of the authors and should not be interpreted as representing the official policies, either expressed or implied, of the Army Research Office or the U.S. Government. The U.S. Government is authorized to reproduce and distribute reprints for Government purposes notwithstanding any copyright notation herein. Additional funding was provided by grant #G2023-20946 from the Alfred P. Sloan Foundation.

## ACKNOWLEDGMENT

The authors thank Roman Saladino for assistance in testing the implementation feasibility in the sample deployment case.